\documentclass[aps,pre,twocolumn,superscriptaddress,showpacs]{revtex4-1}

\usepackage{amssymb}
\usepackage{color}
\usepackage{graphicx}
\usepackage{amsmath}
\usepackage{mathrsfs}
\usepackage{times}
\usepackage{subeqnarray}
\usepackage{cases}
\usepackage{bm}

\begin{document}

\title{Learning Hamiltonian dynamics by reservoir computer}

\author{Han Zhang}
\affiliation{School of Physics and Information Technology, Shaanxi Normal University, Xi'an 710119, China}

\author{Huawei Fan}
\affiliation{School of Physics and Information Technology, Shaanxi Normal University, Xi'an 710119, China}

\author{Liang Wang}
\affiliation{School of Physics and Information Technology, Shaanxi Normal University, Xi'an 710119, China}

\author{Xingang Wang}
\email{Email address: wangxg@snnu.edu.cn}
\affiliation{School of Physics and Information Technology, Shaanxi Normal University, Xi'an 710119, China}

\begin{abstract}
Reconstructing the KAM dynamics diagram of Hamiltonian system from the time series of a limited number of parameters is an outstanding question in nonlinear science, especially when the Hamiltonian governing the system dynamics are unknown. Here, we demonstrate that this question can be addressed by the machine learning approach knowing as reservoir computer (RC). Specifically, we show that without prior knowledge about the Hamilton's equations of motion, the trained RC is able to not only predict the short-term evolution of the system state, but also replicate the long-term ergodic properties of the system dynamics. Furthermore, by the architecture of parameter-aware RC, we also show that the RC trained by the time series acquired at a handful parameters is able to reconstruct the entire KAM dynamics diagram with a high precision by tuning a control parameter externally. The feasibility and efficiency of the learning techniques are demonstrated in two classical nonlinear Hamiltonian systems, namely the double-pendulum oscillator and the standard map. Our study indicates that, as a complex dynamical system, RC is able to learn from data the Hamiltonian. 
\end{abstract}

\maketitle

\section{introduction}\label{intro}

Model-free prediction of chaotic dynamical systems using machine learning approaches has received broad research interest in recent years~\cite{RC:Maass2002,RC:lukosevicius2009,RC:Tanaka2019,RC:Tang2020,RC:Lipton}. In particular, a technique knowing as reservoir computer (RC) has been widely adopted in literature for predicting the state evolution of chaotic systems~\cite{RC:Jaeger,RC:Griffith,RC:Lu,RC:Pathak2018,RC:UP2018,RC:Carroll,RC:Jiang,RF:2019,ML:Cestnik,RC:Fan,JZK:2021}. From the perspective of dynamical systems, RC can be regarded as a complex network of coupled dynamical elements, which, driven by the input data, generates the output data through a readout function. Except the parameters of the readout function, which are to be determined by the training process, all other parameters and the dynamics of RC are fixed at the construction. After training, the system is closed by using the output as the input, and then is evolving as an autonomous system for predictions. Evidences have shown that compared with the conventional prediction techniques in nonlinear science, RC has an obvious advantage in both accuracy and efficiency. For instance, it is demonstrated that a well trained RC can accurately predict the state evolution of chaotic systems for about a half-dozen Lyapunov times, which is much longer than the prediction horizon of conventional techniques~\cite{RC:Jaeger}.

Besides predicting chaos evolutions, RC has also been exploited to address other long-standing questions in nonlinear science, saying, for example, reconstructing chaotic attractors and calculating Lyapunov exponents~\cite{RC:Pathak2017}, synchronizing chaotic oscillators~\cite{RC:Lu,RC:Weng}, predicting system collapses~\cite{KLW:2021}, reconstructing synchronization transition paths~\cite{FHW:arxiv}, transferring knowledge between different systems~\cite{CK:PRL2020,RC:Guo2021}, to name just a few. These studies, while demonstrating the power of RC in solving different nonlinear questions, also give insights on the working mechanisms of RC. For instance, although RC fails to predict the long-term evolution of chaotic systems, it can replicate the ergodic properties of the chaotic systems faithfully, e.g., the Lyapunov exponents and the returned maps~\cite{RC:Pathak2017}. This ability, knowing as climate replication, suggests that it is the intrinsic dynamics of the chaotic system that RC essentially learns from the data, instead of the mathematical expressions describing the time series. Exploiting this ability, model-free techniques have been proposed in recent years to predict the bifurcations in nonlinear dynamical systems, e.g., reproducing the bifurcation diagram of classical chaotic systems~\cite{RF:2019,ML:Cestnik}, anticipating the critical points of system collapse~\cite{KLW:2021}, predicting the critical coupling for synchronization~\cite{FHW:arxiv}, etc. Another property revealed recently in exploiting RC is that knowledge can be transferred between different dynamical systems, namely the ability of transfer learning~\cite{CK:PRL2020,RC:Guo2021}. Specifically, it is shown that the RC trained by the time series of system $A$ can be used to infer the properties of system $B$, with the motions of $A$ and $B$ significantly different from each other. It is worth mentioning that the systems employed in these studies are dominantly dissipative, in the sense that the final dynamics of the system is independent on the initial conditions. The adoption of dissipative systems is nature, as the reservoir network itself can be regarded as a dissipative system. In particular, the memory-fading property of RC requires that the dynamics of the reservoir network should be converged into a low-dimensional manifold independent of the network initial conditions~\cite{RC:lukosevicius2009}.   

From the astronomical scales to the quantum scales, many physical systems are described by the Hamiltonian formalism. Different from dissipative systems, the symplectic structure of the Hamiltonian requires that during the time course of system evolution, the phase-space volume of a closed surface should be preserved and, for time-independent Hamiltonian, the total energy of the system should be conserved~\cite{Book:OTT}. The concerns about volume-preservation and energy-conservation leads to the development of physics-enhanced machine learning techniques in recent years, saying, for instance, the Hamiltonian neural networks (HNN)~\cite{HNNs:Greydanus2019,HNNs:toth2019,HNNs:Bertalan2019,HNNs:Choudhary2020,HNNs:Mattheakis2020,HNNs:Cranmer2020,HNNs:Burby2021,HNNs:AC2021,HNNs:Dulberg2020,HNNs:Lai2021}. The idea of HNN was articulated about three decades ago~\cite{RBL:1993}, where methods for designing neural networks of Hamiltonian dynamics have been proposed. Recent studies generalize this idea by incorporating the Hamiltonian mechanisms into the conventional approaches in machine learning, e.g., deep learning, resulting in the improved learning performance~\cite{HNNs:Greydanus2019,HNNs:toth2019,HNNs:Bertalan2019,HNNs:Choudhary2020,HNNs:Mattheakis2020}. The idea of HNN has been also extended to systems described by Lagrangian formalism and generalized coordinates~\cite{HNNs:Cranmer2020,HNNs:Dulberg2020,HNNs:AC2021}. In particular, a parameter-aware architecture has been proposed for reconstructing the Kolmogorov-Arnold-Moser (KAM) dynamics diagram of Hamiltonian systems, in which the adaptability of HNN has been explored~\cite{HNNs:Lai2021}. It is worth mentioning that while HNN has been proven as efficient for learning Hamiltonian dynamics, the training of HNN is time-consuming as compared with RC. For instance, due to the multiple-layered network structure, the number of parameters to be trained in HNN is orders of magnitude larger to that of RC. In addition, in designing HNN, physics constraints (e.g., the Hamilton's equations of motion) are enforced to the machine, yet it remains not clear whether such an enforcement is necessary, neither do we understand completely the mechanism underlying the improved performance.

From the standpoint of state evolution prediction, the mission of RC and HNN is very similar, i.e., mimicking the function relating the input and output. Whereas the output of RC does not satisfy the symplectic properties of the Hamiltonian, a high-dimensional reservoir does has the ability to fit any form of function by a property training of the output matrix~\cite{RC:Maass2002}. Once the statistical properties of the time series are captured by the output of RC, the climate of the Hamiltonian dynamics will be replicated. Furthermore, using the property of transfer learning, we might also able to reconstruct the KAM dynamics diagram by the RC trained at a limited number of parameters. If it is indeed the case, the computational cost as required by HNN will be significantly reduced. Our main objective in the present work is just to verify this speculation. To be specific, by a recently proposed architecture of parameter-aware RC~\cite{KLW:2021,FHW:arxiv}, we train RC by the time series of Hamiltonian systems acquired at a handful of parameters, and then use the trained RC to replicate the dynamics climates. We are able to show that the trained RC not only is able to replicate the climates associated with the training parameters, but also is able to generate the entire KAM dynamics diagram with a high precision by tuning a control parameter externally.  

In the following section, we will present the architecture of parameter-aware RC used for predicting Hamiltonian dynamics, together with the training method. The application of RC to the prediction of Hamiltonian dynamics and the reconstruction of the KAM for two classical Hamiltonian systems, namely the double-pendulum oscillator and the standard map, will be reported in Sec. III. Discussions and conclusion will be given in Sec. IV.

\section{parameter-aware reservoir computer}\label{rc}

We adopt the architecture of parameter-aware RC to learn the dynamics of Hamiltonian systems~\cite{KLW:2021,FHW:arxiv}. In this architecture, the RC is constituted by four modules: the $I/R$ layer (input-to-reservoir), the parameter-control module, the reservoir network, and the $R/O$ layer (reservoir-to-output). The $I/R$ layer is characterized by the matrix $\mathbf{W}_{in}\in\mathbb{R}^{D_r\times D_{in}}$, which couples the input vector $\mathbf{u}_{\beta}(t)\in\mathbb{R}^{D_{in}}$ to the reservoir network. Here, $\mathbf{u}_{\beta}(t)$ denotes the input vector that is acquired from the system at time $t$ and under the specific system parameter $\beta$. The elements of $\mathbf {W}_{in}$ are randomly drawn from a uniform distribution within the range $[-\sigma, \sigma]$. The parameter-control module is characterized by the vector $\mathbf{s}=\beta\mathbf{b}$, with $\beta$ the control parameter and $\mathbf{b}\in \mathbb{R}^{D_{r}}$ the bias vector. In applications, the control parameter $\beta$ can be regarded as an additional input marking the input vector $\mathbf{u}(t)$. In our studies, we choose $\beta$ to be the initial conditions of the Hamiltonian system, as the variation of which can lead to different dynamical motions. The elements of $\mathbf{b}$ are drawn randomly from a uniform distribution within the range $[-\sigma, \sigma]$. The reservoir network contains $D_r$ dynamical nodes, with the initial states of the nodes being randomly chosen from the interval $[-1,1]$. The states of the nodes in the reservoir network, $\mathbf{r}(t)\in \mathbb{R}^{D_r}$, are updated according to the equation
\begin{equation}\label{rc1}
\mathbf{r}(t+\Delta t)=(1-\alpha)\mathbf{r}(t)+\alpha\tanh[\mathbf {A}\mathbf{r}(t)+\mathbf{W}_{in}\mathbf{u}_{\beta}(t)+\beta\mathbf{b}].
\end{equation}
Here, $\Delta t$ is the time step for updating the reservoir, $\alpha$ is the leaking coefficient, and $\mathbf{A}\in \mathbb{R}^{D_r\times D_r}$ is the weighted adjacency matrix representing the coupling relationship between nodes in the reservoir. The adjacency matrix $\mathbf{A}$ is constructed as a sparse random Erd\"{o}s-R\'{e}nyi network: with the probability $d$, each element of the matrix is arranged a nonzero value drawn randomly from the interval $[-1,1]$. The matrix $\mathbf{A}$ is rescaled, so as to make its spectral radius equal $\rho$. The output layer is characterized by the matrix $\mathbf{W}_{out}\in \mathbb{R}^{D_{out}\times D_{r}}$, which generates the output vector, $\mathbf{v}(t)\in \mathbb{R}^{D_{out}}$, by the operation
\begin{equation}\label{rc2}
\mathbf{v}(t+\Delta t)=\mathbf{W}_{out}\mathbf{r}(t+\Delta t),
\end{equation}
with $\mathbf{W}_{out}$ the output weight matrix to be estimated through the training process. Except $\mathbf{W}_{out}$, all other parameters of the RC, e.g., $\mathbf{W}_{in}$ and $\mathbf{A}$, are fixed at the construction. Briefly, the purpose of the training process is to find a suitable output matrix $\mathbf{W}_{out}$ so that the output vector $\mathbf{v}(t+\Delta t)$ as calculated by Eq. (\ref{rc2}) is as close as possible to the input vector $\mathbf{u}(t+\Delta t)$ for $t=(\tau+1)\Delta t,\ldots,(\tau+L)\Delta t$, with $T_0=\tau\Delta t$ the transient period discarded in data acquisition and $L$ the length of the training time series. This can be done by minimizing the following the cost function with respect to $\mathbf{W}_{out}$~\cite{RC:Pathak2017,RC:Lu,RC:Pathak2018}
\begin{equation}\label{rc3}
\mathbf{W}_{out}=\mathbf{U}\mathbf{V}^T(\mathbf{V}\mathbf{V}^T+\lambda\mathbb{I})^{-1}.
\end{equation}
Here, $\mathbf{V}\in \mathbb{R}^{D_{r}\times L}$ is the state matrix whose $k$th column is $\mathbf{r}[(\tau+k)\Delta t]$, $\mathbf{U}\in \mathbb{R}^{D_{in}\times L}$ is a matrix whose $k$th column is $\mathbf{u}[(\tau+k)\Delta t]$, $\mathbb{I}$ is the identity matrix, and $\lambda$ is the ridge regression parameter for avoiding the overfitting. After training, the output matrix $\mathbf{W}_{out}$ will be fixed, and the RC is ready for prediction. In the predicting phase, first we set the control parameter $\beta$ to a specific value of interest (not necessarily the parameters used in the training phase), and then evolve the RC as an autonomous dynamical system by taking the output vector $\mathbf{v}(t)$ as the next input vector $\mathbf{u}_{\beta}(t)$. Finally, by a fine tuning of the control parameter $\beta$, we try to reconstruct the KAM diagram of the system.

We note that the input data in the training phase contains two time series: (1) the input vector $\mathbf{u}_{\beta}(t)$ that represents the system state and (2) the control parameter $\beta(t)$ that labels the condition under which the input vector $\mathbf{u}_{\beta}(t)$ is acquired. More specifically, the input vector $\mathbf{u}_{\beta}(t)$ is composed of $m$ segments of length $T$, while each segment is a time series obtained under a specific control parameter $\beta$. As such, $\beta(t)$ is a step-function of time. In the predicting phase, besides replacing $\mathbf{u}_{\beta}(t)$ with $\mathbf{v}(t)$, we still need to input the control parameter $\beta(t)$, so as to guide the reservoir evolution. For convenience, we set $D_{in}=D_{out}$ for the input and output vectors. 

While there are many choices for the control parameter in Hamiltonian systems, we choose the initial conditions. To be specific, we generate different motions by varying one of the initial conditions, while keeping the other initial conditions unchanged. We are going to show that the parameter-aware RC is able to not only replicate the dynamics climates associated with the training parameters (initial conditions), but also reproduce the dynamics climates associated with other parameters (initial conditions), thereby reconstructing the entire dynamics diagram of the Hamiltonian system.

\section{Results}

\subsection{The double-pendulum oscillator: results for standard RC}

We start by showing the capability of the standard RC in predicting and replicating the dynamics of Hamiltonian systems. As discussed in the above section, a common sense in the existing studies of machine learning is that RC is applicable to only dissipative dynamical systems, and, to predict and replicate Hamiltonian systems, physics-constraints from the Hamiltonian mechanisms should be incorporated into the learning algorithm, e.g., the development of HNN~\cite{HNNs:Greydanus2019,HNNs:toth2019,HNNs:Bertalan2019,HNNs:Choudhary2020,HNNs:Mattheakis2020}. Therefore, before reconstructing the KAM diagram by the method of parameter-aware RC, we need to check first whether the dynamics of Hamiltonian systems can be learned by the standard RC, which is realized by setting $\mathbf{b}=0$ in Eq.~(\ref{rc1}) and $m=1$ in preparing the training data. 

The first model of Hamiltonian system investigated in our study is the double-pendulum oscillator. Double-pendulum is a classical model in textbook for demonstrating the nonlinear dynamics of Hamiltonian systems~\cite{RBL:1993,double2020}. The Hamiltonian of double-pendulum oscillator is $\mathcal{H}=E_k+E_p=[(m_1/6+m_2/2)l_1^2\omega_1^2+m_2l_2^2\omega_2^2/6+m_2l_2l_1\omega_1\omega_2\cos(\theta_1-\theta_2)/2]-g[(m_1/2+m_2)l_1\cos\theta_1+l_2m_2\cos\theta_2/2]$, with $\omega_{1,2}=d\theta_{1,2}/dt$ the angular frequencies and $g=9.8 \ m/s^2$ the acceleration of gravity on Earth. The variables $\theta_{1,2}$, $m_{1,2}$ and $l_{1,2}$ denote the angular displacements, masses and lengths of the two pendulums, respectively. By changing the initial values of the two pendulums, $\mathbf{u}(0)=[\theta_1(0),\omega_1(0),\theta_2(0),\omega_2(0)]$, the system can present rich dynamical behaviors, including quasi-periodic and chaotic motions. According to Lagrange's equation of the second kind, the dynamics of the double-pendulum oscillator is governed by equations
\begin{widetext}
\begin{eqnarray}
\begin{split}
(\frac{m_{1}}{3}+m_{2})l_{1}^2\dot{\omega}_{1}+\frac{m_{2}l_{1}l_{2}}{2}\cos(\theta_{1}-\theta_{2})\dot{\omega}_{2}+\frac{m_{2}l_{1}l_{2}}{2}\sin(\theta_{1}-\theta_{2})\omega_{2}^2+\frac{(m_{1}+2m_{2})gl_{1}}{2}\sin\theta_{1}=0&,\\ 
\frac{m_{2}l_{1}l_{2}}{2}\cos(\theta_{1}-\theta_{2})\dot{\omega}_{1}+\frac{m_{2}l_{2}^2}{3}\dot{\omega}_{2}-\frac{m_{2}l_{1}l_{2}}{2}\sin(\theta_{1}-\theta_{2})\omega_{1}^2+\frac{m_{2}gl_{2}}{2}\sin\theta_{2}=0&.
\end{split}
\end{eqnarray}
\end{widetext}
Without the loss of generality, we set the two pendulums to be identical in mass and length, i.e., $m_1=m_2=m$ and $l_1=l_2=l$. By introducing the new time variable $t=\sqrt{g/l_{1}}\tau$ [$\tau$ is time value for Eq. (4)], the equations of motion of the double-pendulum oscillator can be rewritten as
\begin{widetext}
\begin{eqnarray}
\begin{split}
\label{ode1}
&\dot{\omega}_{1}=[9\cos(\theta_{1}-\theta_{2})\sin(\theta_{1}-\theta_{2})\omega_{1}^2+6\sin(\theta_{1}-\theta_{2})\omega_{2}^2+18\sin\theta_{1}-9\cos(\theta_{1}-\theta_{2})\sin\theta_{2}]/[9\cos^2(\theta_{1}-\theta_{2})-16],\\
&\dot{\omega}_{2}=[24\sin(\theta_{1}-\theta_{2})\omega_{1}^2+9\cos(\theta_{1}-\theta_{2})\sin(\theta_{1}-\theta_{2})\omega_{2}^2+27\cos(\theta_{1}-\theta_{2})\sin\theta_{1}-24\sin\theta_{2}]/[16-9\cos^2(\theta_{1}-\theta_{2})].
\end{split}
\end{eqnarray}
\end{widetext}
The total energy of the system now reads $E=2\omega_1^2/3+\omega_2^2/6+[\omega_1\omega_2\cos(\theta_1-\theta_2)-\cos(\theta_2)-3\cos(\theta_1)]/2$. In simulations, Eq. (5) is evolved numerically by the symplectic algorithm, with time step being set as $\Delta t=0.2$. In our studies, we fix the initial conditions $\theta_{1}(0)=0.6$, $\omega_{1}(0)=0$, $\omega_{2}(0)=0$, while changing the initial condition $\theta_{2}(0)$ within the range $[-\pi,\pi)$ to generate different motions. 

\begin{figure*}[tbp]
\begin{center}
\includegraphics[width=0.85\linewidth]{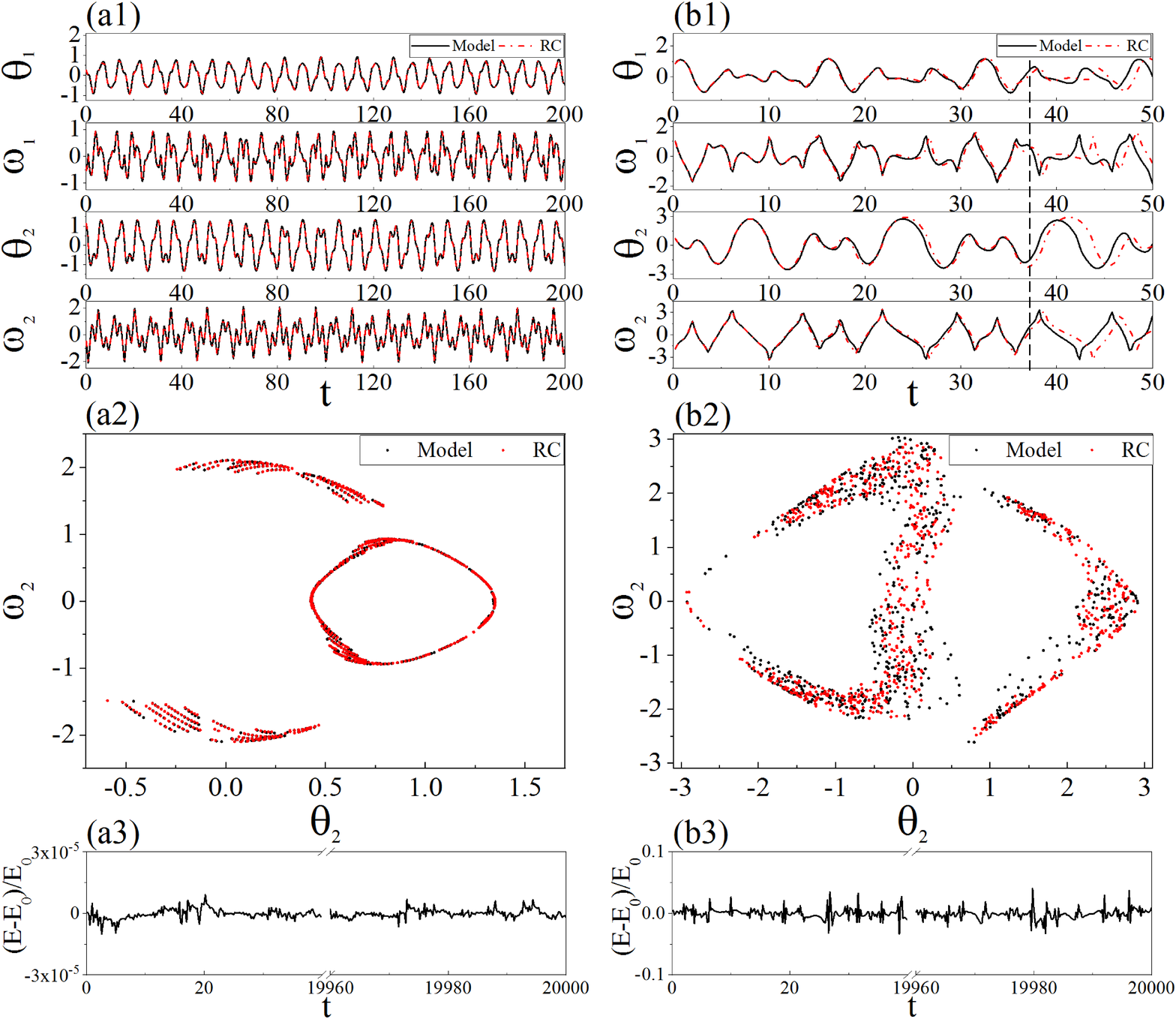}
\caption{Predicting the dynamics of double-pendulum oscillator by the standard RC. (a) Predicting the quasi-periodic motion generated by the initial condition $\theta_{2}(0)=1.35$. (a1) The time evolution of the system state. (a2) The system trajectory detected by the Poincar\'{e} surface of section defined by $\omega_1=0$ and $\theta_1>0$. (a3) The long-time evolution of the system energy. The true energy of the system is $E_0=-1.3475$. (b) Predicting the chaotic motion generated by the initial condition $\theta_{2}(0)=2.04$. (b1) The time evolution of the system state. (b2) The chaotic trajectories detected by the Poincar\'{e} surface of section. (b3) The long-time evolution of the system energy. The true energy of the system is $E_0=-1.01$. Results obtained by the direct simulations of the model system are colored in black. Results predicted by RC are colored in red.}
\label{fig1}
\end{center}
\end{figure*}

We demonstrate first the learning of an integrable Hamiltonian system. Setting $\theta_{2}(0)=1.35$, the oscillator presents the quasi-periodic motion, as depicted in Fig.~\ref{fig1}(a) (the black curves). By solving Eq. (5) numerically, we collect the system state $\mathbf{u}(t)=[\theta_{1}(t), \omega_{1}(t), \theta_{2}(t), \omega_{2}(t)]^T$ for a sequence of $\hat{T}=3\times10^3$ time steps (about $100$ oscillation cycles). The sequence is divided into two segments. The first segment of length $T=2\times 10^3$ is used as the training data, and the second segment of length $T'=1\times 10^3$ is used as the test data. The training data is then used to calculate the output matrix $\mathbf{W}_{out}$ of the RC. In this case, the parameters of the reservoir are chosen as $(D_{r}, d, \rho, \alpha, \sigma, \lambda)=(500, 0.48, 1.48, 0.25, 1.52, 1\times10^{-9})$, which are obtained by the optimizer ``optimoptions" in MATHLAB. In the predicting phase, the final state of the reservoir network in the training phase is used as the initial state, and the reservoir is evolving according to Eqs. (1) and (2) by replacing $\mathbf{u}(t)$ with $\mathbf{v}(t)$. The time evolution of the system state predicted by the trained RC is plotted in Fig.~\ref{fig1}(a1) (the red curves). We see that the predictions are in good agreement with the results obtained from direct simulations of the model system for a long period. To check whether the statistical properties of the quasi-periodic motion is properly replicated by RC, we plot in Fig.~\ref{fig1}(a2) the system trajectory on the Poincar\'{e} surface of section defined by $\omega_1=0$ and $\theta_1>0$. We see that the trajectory predicted by the RC is well overlapped with the one obtained from the model system, manifesting the proper replication of the dynamics climate. The results in Figs.~\ref{fig1}(a1) and (a2) validate the capability of the standard RC in predicting and replicating the dynamics of integrable Hamiltonian system.

We next demonstrate the learning of non-integrable Hamiltonian system. Setting $\theta_{2}(0)=2.04$, the oscillator presents the chaotic motion, with the largest Lyapunov exponent being $\Lambda\approx 0.163$. In the similar way, we generate the training data by simulating Eqs. (1) and (2), calculate the output matrix $\mathbf{W}_{out}$, and then use the trained RC to predict and replicate the system dynamics. For the case of Hamiltonian chaos, the set of parameters for the RC are $(D_{r}, d, \rho, \alpha, \sigma, \lambda)=(500, 0.36, 2.66, 0.24, 2.08, 5.4\times10^{-2})$. Fig.~\ref{fig1}(b1) shows the time evolution of the system state predicted by RC (the red curves), together with the results obtained from model simulations (the black curves). We see that the RC can predict accurately the system evolution for a period of $T\approx 35$ (about $6$ Lyapunov times). Fig.~\ref{fig1}(b2) shows the system trajectory on the Poincar\'{e} surface of section defined by $\omega_1=0$ and $\theta_1>0$. We see that the predicted and true trajectories are well overlapped in the space, indicating that the climate of Hamiltonian chaos is successfully replicated. To confirm further the replication of the system climate, we calculate the largest Lyapunov exponent of the predicted trajectory by the numerical method proposed in Ref.~\cite{MS:1985}. The calculated result is $\Lambda\approx 0.166$, which agrees with the one obtained from direct simulations well. 

In machine learning of Hamiltonian systems, a major concern is whether the system energy will be conserved in the long-term evolution~\cite{HNNs:Greydanus2019,HNNs:Cranmer2020}. In particular, it has been shown that if the physics constraints of the Hamiltonian and Lagrangian mechanisms are not imposed in the feedforward neural network, the system energy predicted by the machine will be gradually decreased as time increases. This concern has led to the development of HNN and Lagrangian neural networks (LNN)~\cite{HNNs:Greydanus2019,HNNs:Cranmer2020}, with which the system energy is well conserved in the long-term evolution. The problem of energy conservation, however, remains as an open issue for RC, as the architecture of RC is completely different from that of the feedforward neural networks. The dissipative nature of the reservoir suggests that the system energy predicted by the machine could be decreasing with time, whereas the results of dynamics prediction and climate replication in Fig.~\ref{fig1} suggest the conservation of the system energy. To check out, we plot in Figs.~\ref{fig1}(a3) and (b3) the long-time evolution of the system energy for the quasi-periodic and chaotic motions, resepctively. We see that, surprisingly, the system energy is well conserved for both cases. For the quasi-periodic motion [see Fig.~\ref{fig1}(a3)], the system energy is fluctuating around the true value $E_0$ by small amplitudes of the order of $10^{-5}$; for the chaotic motion [see Fig.~\ref{fig1}(b3)], the fluctuating amplitudes are of the order of $10^{-2}$. We note that, by decreasing the time step $\Delta t$, the fluctuating amplitudes can be further decreased (not shown).  

\begin{figure}[tbp]
\begin{center}
\includegraphics[width=0.8\linewidth]{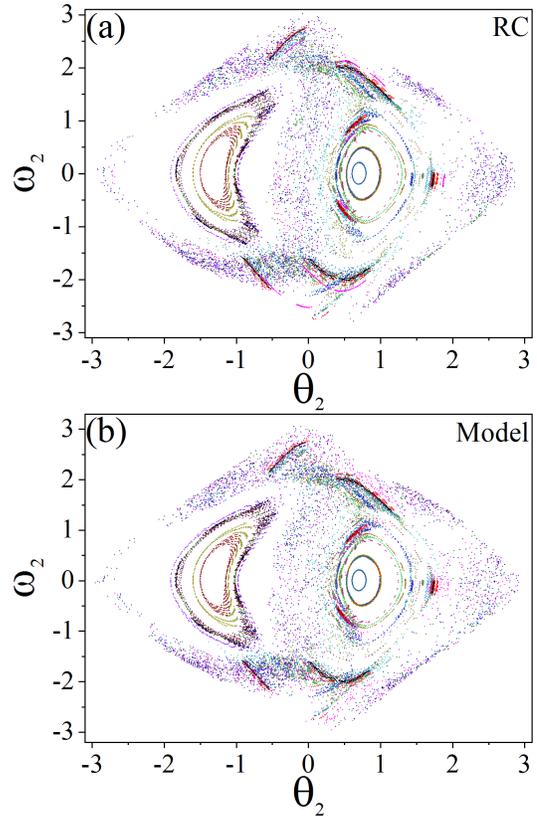}
\caption{The dynamics of $n=34$ different double-pendulum oscillators on the Poincar\'{e} surface of section. (a) The results predicted by the standard RC. (b) The results from direction simulations of the model system. In (a) and (b), trajectories with the same initial value $\theta_2(0)$ are represented by the same color.}
\label{fig2}
\end{center}
\end{figure}

To check the generality of the standard RC in replicating the climate of Hamiltonian dynamics, we keep the other initial values of the double-pendulum oscillator unchanged, while choosing the initial value of $\theta_2(0)$ randomly within the range $[-\pi,\pi)$. As above, for each value of $\theta_2(0)$, we first train the RC by the time series obtained from model simulations, and then replicate the dynamics climate based on the RC outputs. $n=34$ initial values are chosen in total. The trajectories of the replicated dynamics on the Poincar\'{e} surface of section ($\omega_1=0$ and $d\omega_1/dt>0$) are plotted in Fig.~\ref{fig2}(a). We see that the reconstructed KAM diagram is mixed with chaotic and regular dynamics. By the same set of initial values of $\theta_2(0)$, we plot in Fig.~\ref{fig2}(b) the KAM diagram based on the results of model simulations. Comparing Fig.~\ref{fig2}(a) with Fig.~\ref{fig2}(b), we see that the true KAM diagram is well replicated by RC.

\subsection{The double-pendulum oscillator: results for parameter-aware RC}\label{dpob}

We move on to study the capability of parameter-aware RC in replicating the KAM dynamics diagram, which now employs the parameter-control module (i.e. the vector $\beta \mathbf{b}$) in Eq. (\ref{rc1})~\cite{KLW:2021,FHW:arxiv,HNNs:Lai2021}. The adoption of the parameter-aware RC is motivated by the fact that in many realistic situations only the time series of a limited number of system motions are available, while the mission is to predict the dynamics, or replicate the system climate, of many unknown motions. As the knowledge the machine learned from the training system (e.g. the output matrix) is used to predict the dynamics of other systems, parameter-aware RC thus can be regarded as another approach of transfer learning~\cite{TL:SJP2010,TL:FZ2015}. For the case of Hamiltonian dynamics, our mission here is replicating the KAM dynamics diagram (as shown in Fig.~\ref{fig2}) based on the time series of a handful of system motions. 

\begin{figure}[tbp]
\begin{center}
\includegraphics[width=0.85\linewidth]{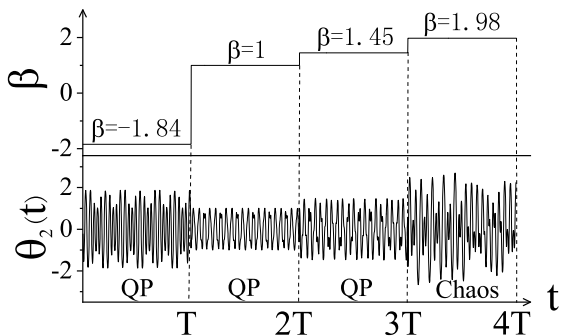}
\caption{For the double-pendulum oscillator, a schematic plot of the training data used in training the parameter-aware RC. The training data consists of $m=4$ segments of equal length $T=2.5\times10^{3}$. The segments are generated by the control parameters $\beta=-1.84$, $1$, $1.45$ and $1.98$. For $\beta=-1.84$, $1$, and $1.45$, the system dynamics is quasi-periodic (QP); for $\beta=1.98$, the system dynamics is chaotic.}
\label{fig3}
\end{center}
\end{figure}

Different from the standard RC, in parameter-aware RC the training data is composed of $m$ segments, with each segment being a time series of length $T$ generated under a specific control parameter $\beta$. To generate the training data in simulations, we fix again the initial values $[\theta_{1}(0),\omega_{1}(0),\omega_{2}(0)]=(0.6,0,0)$, while changing $\theta_{2}(0)$ to adjust the system dynamics. In this application, we set $\theta_2(0)$ as the control parameter, i.e., setting $\beta=\theta_2(0)$ in Eq. (\ref{rc1}). The $m$ segments are then combined to form the new time series $\mathbf{u}_{\beta}(t)=[\theta_{1}(t), \omega_{1}(t), \theta_{2}(t), \omega_{2}(t)]_{\beta}$, which is of length $L=mT$. The new time series and the time series of the control parameter $\beta(t)$ are fed into the reservoir for estimating the output matrix $\mathbf{W}_{out}$. As an illustration, we choose $m=4$ control parameters and acquire for each parameter a time series of $T=2.5\times 10^3$ states. The $4$ control parameters are $\beta=-1.84$, $1.0$, $1.45$, and $1.98$, which are randomly chosen within the range $[-\pi,\pi)$. The system dynamics is quasi-periodic for $\beta=-1.84$, $1.0$ and $1.45$, and is chaotic for $\beta=1.98$ (the largest Lyapunov exponent is $\Lambda\approx 0.16$). The structure of the training data is schematically shown in Fig.~\ref{fig3}. In this application, the set of parameters for the parameter-aware RC are chosen as $(D_{r}, d, \rho, \alpha, \sigma, \lambda)=(1\times 10^3, 0.97, 1.13, 0.64,0.94, 2\times10^{-2})$.

We check first the feasibility of the trained RC in replicating the dynamics climates associated with the training parameters. This is implemented by changing the control parameter $\beta$ to one of the training parameters, and then evolving the reservoir according to Eqs. (1) and (2) as described in Sec. II. We note that the major difference between the standard RC and the parameter-aware RC lies in the variability of the output matrix in the predicting phase. For the standard RC employed in Sec. III A, the output matrix is trained separately for each time series, and each output matrix is only able to replicate the dynamics climate associated to a specific control parameter. For the parameter-aware RC, the output matrix is trained only once, and in the predicting phase the same output matrix is used to reproduce the dynamics climate of any desire parameter. Setting $\beta=-1.84$, we plot in Fig.~\ref{fig3}(a1) the state evolution of the system predicted by the machine, together with the results obtained from model simulations. We see that the state evolution is well predicted by the machine. Fig.~\ref{fig3}(a2) shows the dynamics on the Poincar\'{e} surface of section. We see that dynamics climate is also properly replicated by the RC. The results for $\beta=1.98$ are shown in Fig.~\ref{fig3}(b). We see that the RC is able to predict the evolution for about $6$ Lyapunov times [see Fig.~\ref{fig3}(b1)] and the dynamics climate replicated by the RC is well overlapped with the true one [see Fig.~\ref{fig3}(b2)]. The similar results are also observed for other two training parameters, $\beta=1$ and $1.45$ (not shown). 

\begin{figure}[tbp]
\centering
\includegraphics[width=\linewidth]{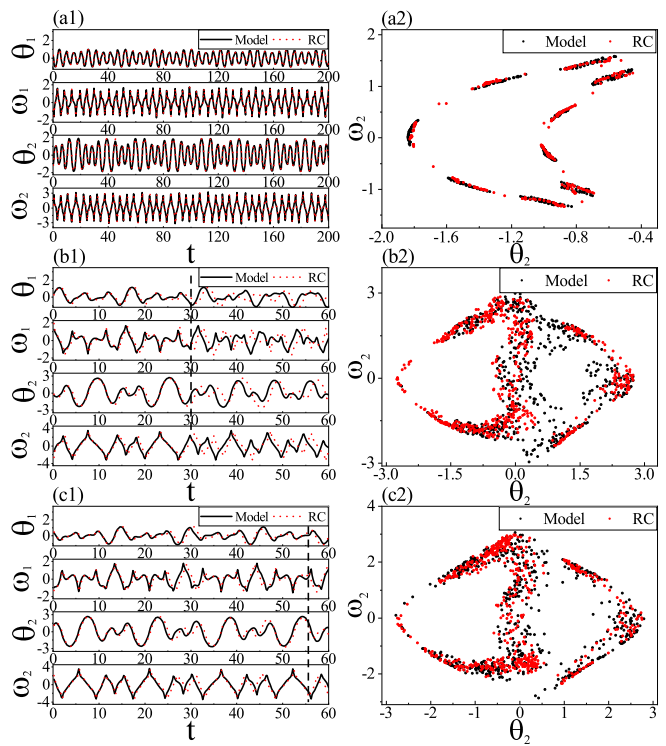}
\caption{Predicting the state evolution and replicating the climate of double-pendulum oscillator by parameter-aware RC. (a) The results for the control parameter $\beta=-1.84$, which is one of the training parameters. (b) The results for the control parameter $\beta=1.98$, which is also one of the training parameters. (c) The results for the control parameter $\beta=2.0$, which is not a training parameter. The vertical lines in (b1) and (c1) denote the prediction horizons. In all graphs, the results predicted by RC are shown as red, and the results obtained from model simulations are shown as black.}
\label{fig4}
\end{figure}

We check next the capability of the trained RC in replicating the dynamics climate of a new parameter not in the training set. As the demonstration, we choose $\beta=2.0$, with which the model system shows chaotic motion and the largest Lyapunov exponent is $\Lambda\approx 0.1$. The results for this new parameter are plotted in Fig.~\ref{fig3}(c). We see that the machine not only predicts accurately the short-time evolution of the system [see Fig.~\ref{fig3}(c1)], but also reproduces properly the system climate [see Fig.~\ref{fig3}(c2)].

\begin{figure}[tbp]
\centering
\includegraphics[width=0.8\linewidth]{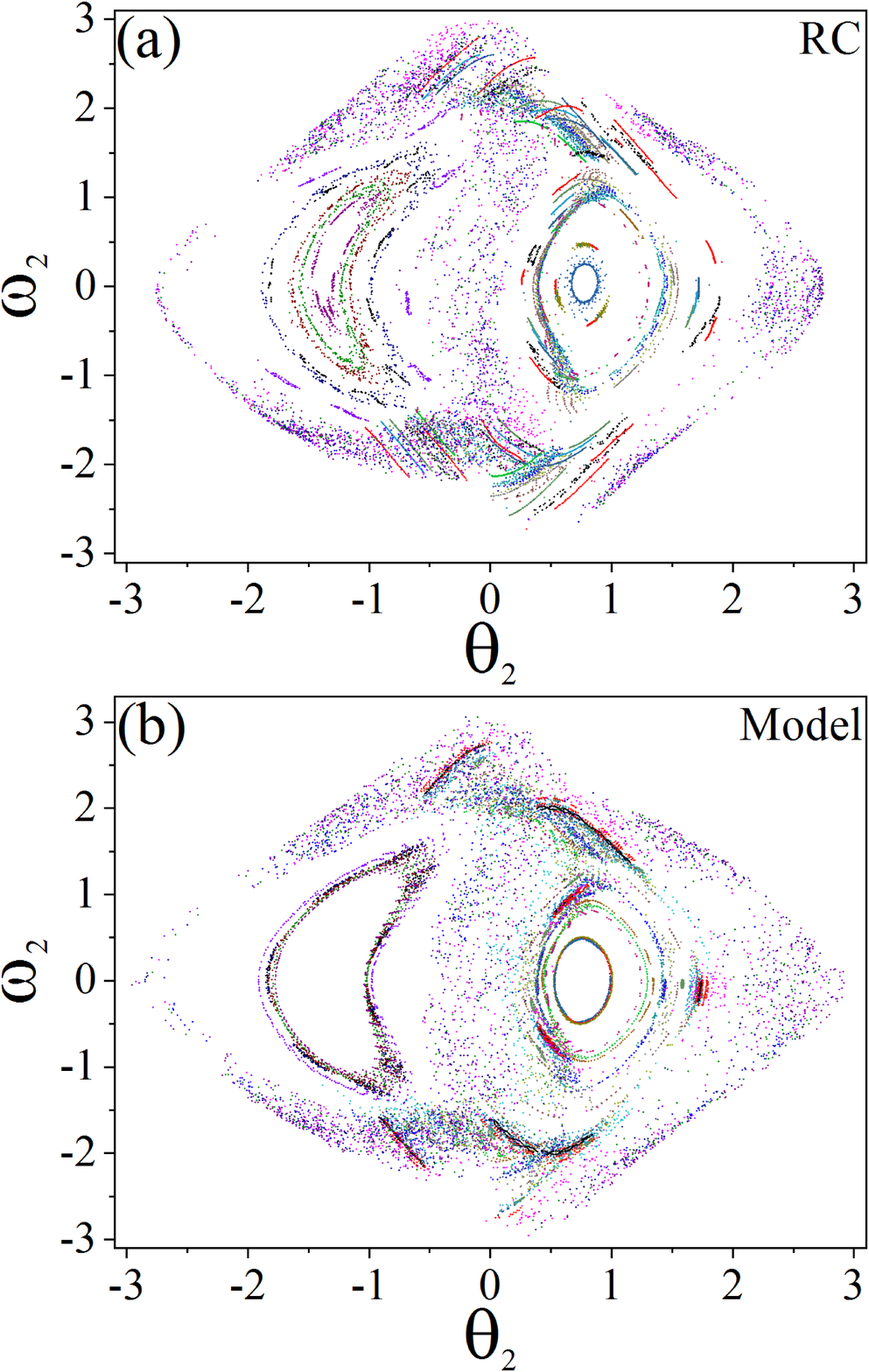}
\caption{The KAM diagram of the double-pendulum oscillator for $n=31$ dynamics. (a) The results predicted by the parameter-aware RC. (b) The results obtained from model simulations. The same set of control parameters (initial conditions) are used in (a) and (b).}
\label{fig5}
\end{figure}

Having justified the capability of parameter-aware RC in replicating the climates of both the training and non-training parameters, we finally exploit it to reconstruct the entire KAM dynamics diagram. In doing this, we keep the output matrix $\mathbf{W}_{out}$ unchanged, while changing the control parameter $\beta$ to a number of values that are randomly chosen within the range $[-\pi,\pi)$. For each value of $\beta$, we collect the output of the machine for $T=1\times 10^4$ steps, based on which we reconstruct the dynamics climate for this specific control parameter. To demonstrate, we choose $n=31$ control parameters. The $n=31$ dynamics on the Poincar\'{e} surface of section are plotted in Fig.~\ref{fig5}(a). We see that the diagram is mixed with quasi-periodic and chaotic motions. By the same set of control parameters, we plot in Fig.~\ref{fig5}(b) the KAM dynamics diagram based on the results of model simulations. Comparing Fig.~\ref{fig5}(a) with Fig.~\ref{fig5}(b), we see that the KAM diagram is well reconstructed by the machine.

\subsection{The standard map}\label{sm}

We generalize the results by reconstructing the KAM diagram of Hamiltonian mapping systems. The model we adopt is the standard map (also known as the kicked rotor). The Hamiltonian of the system reads $\mathcal{H}=p_{\theta}^2/(2I)+K\cos\theta\sum_{n}\delta(t-n\tau)$, with $I$ the rotational inertia of the bar, $K$ the kicking strength, $n\tau$ the moment of the $n$th kick, and $\delta(\ldots)$ the Dirac delta function. $\theta$ and $p_{\theta}$ represent the angular displacement and angular momentum of the bar, respectively. The Hamilton's equations of motion read
\begin{eqnarray}
dp_{\theta}/dt&=&K\sin\theta\sum_n\delta(t-n\tau), \\
d\theta/dt&=&p_{\theta}/I.
\end{eqnarray}   
The integration of the above equations from $t=n\tau$ to $t=(n+1)\tau$ leads to the new equations
\begin{eqnarray}
\theta_{n+1}&=&(\theta_{n}+p_{n})\quad \text{modulo}\quad 2\pi, \\
p_{n+1}&=&(p_{n}+K\theta_{n+1}) \quad \text{modulo}\quad 2\pi.
\end{eqnarray}     
For the sake of simplicity, here we set $\tau=I$, and restrict $\theta$ and $p$ to be within the range $[0,2\pi)$. The KAM dynamics diagram of the standard map is dependent on the kicking strength $K$. When $K$ is small, the diagram is dominated by quasi-periodic dynamics. As $K$ increases, more tori will be broken and the diagram is mixed with quasi-periodic and chaotic dynamics. When $K$ is large, most of the tori will be broken and the diagram is dominated by chaotic motions. Our mission here is to reconstruct the KAM diagram fora fixed value of $K$ based on the information of a handful of motions. 

\begin{figure*}[tbp]
\centering
\includegraphics[width=0.8\linewidth]{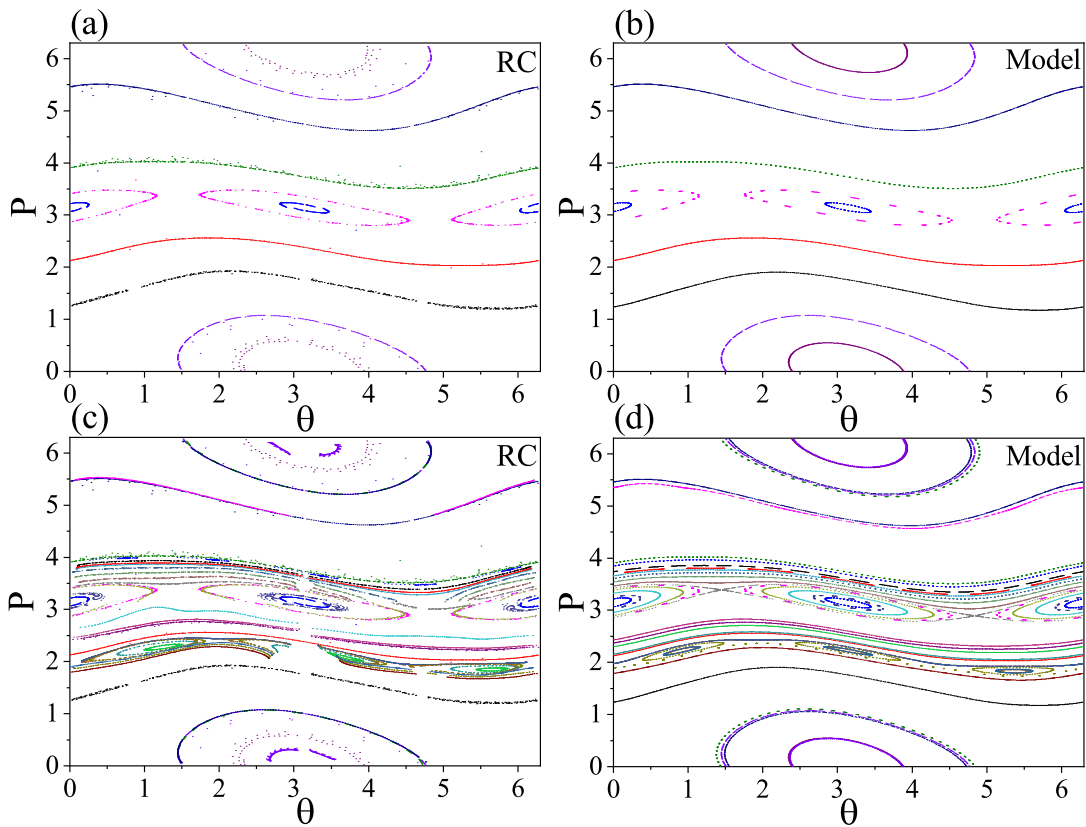}
\caption{Reconstructing the KAM diagram of the standard map under the kicking strength $K=0.5$. The training data is generated by $m=8$ training parameters. (a) The dynamics predicted by the machine for the training parameters. (b) The dynamics of the training parameters obtained from model simulations. (c) The diagram reconstructed by the machine, which includes $m=8$ dynamics from the training parameters and $n=26$ dynamics from the additional control parameters. (d) For the same set of control parameters used in (c), the diagram obtained from model simulations.}
\label{fig6}
\end{figure*}

We demonstrate first the reconstruction of the KAM diagram for a small kicking strength. For demonstration purpose, we choose $K=0.5$, by which the KAM diagram is dominated by quasi-periodic motions. In generating the training data, we fix the initial value of the angular displacement as $\theta_0=\pi$, while changing the initial value of the angular momentum $p_0$ to $m=8$ different values. The $8$ initial values of $p_0$ are randomly chosen within the range $[0,2\pi)$, which are $1.76, 2.38,3.2,3.35,3.73,4.74,5.28$ and $5.77$ in this case. For each value of $p_0$, we simulate the system dynamics according to Eqs. (8) and (9), and record the system state, $\mathbf{u}(n)=[\sin(\theta_{n}), \sin(p_{n}), \cos(\theta_{n}), \cos(p_{n})]^T$ (which is generalized from the state $[\theta_{n}, p_{n}]^T$), for a time series of $T=2\times 10^3$ iterations. The training data therefore is of length $L=mT=1.6\times 10^4$, which, together with the corresponding time series of the control parameter ($\beta=p_0/2\pi$), are fed into the parameter-aware RC for estimating the output matrix. In the predicting phase, we keep the output matrix fixed, while tuning the control parameter to different values and, based on the predictions, reconstruct the KAM diagram. In this application, the parameters of the reservoir are $(D_{r}, d, \rho, \alpha, \sigma, \lambda)=(1.5\times 10^3, 3.6\times 10^{-3}, 1.62, 0.95, 1.59, 8.2\times10^{-2})$. Figure~\ref{fig6}(a) shows the dynamics climates predicted by the machine for the $8$ training parameters. The corresponding dynamics obtained from direct simulations of the model system are plotted in Fig.~\ref{fig6}(b). We see that the predictions are in good agreement with the true results. To reconstruct the entire KAM diagram, we change the control parameter $\beta$ to $n=26$ new values that are randomly chosen within the range $(0,1)$. The dynamics climates of the $n=26$ new parameters, together with the ones of the $8$ training parameters, are plotted in Fig.~\ref{fig6}(c). The corresponding results obtained from model simulations are plotted in Fig.~\ref{fig6}(d). We see that the KAM diagram is well replicated by the machine.  

\begin{figure*}[tbp]
\centering
\includegraphics[width=0.8\linewidth]{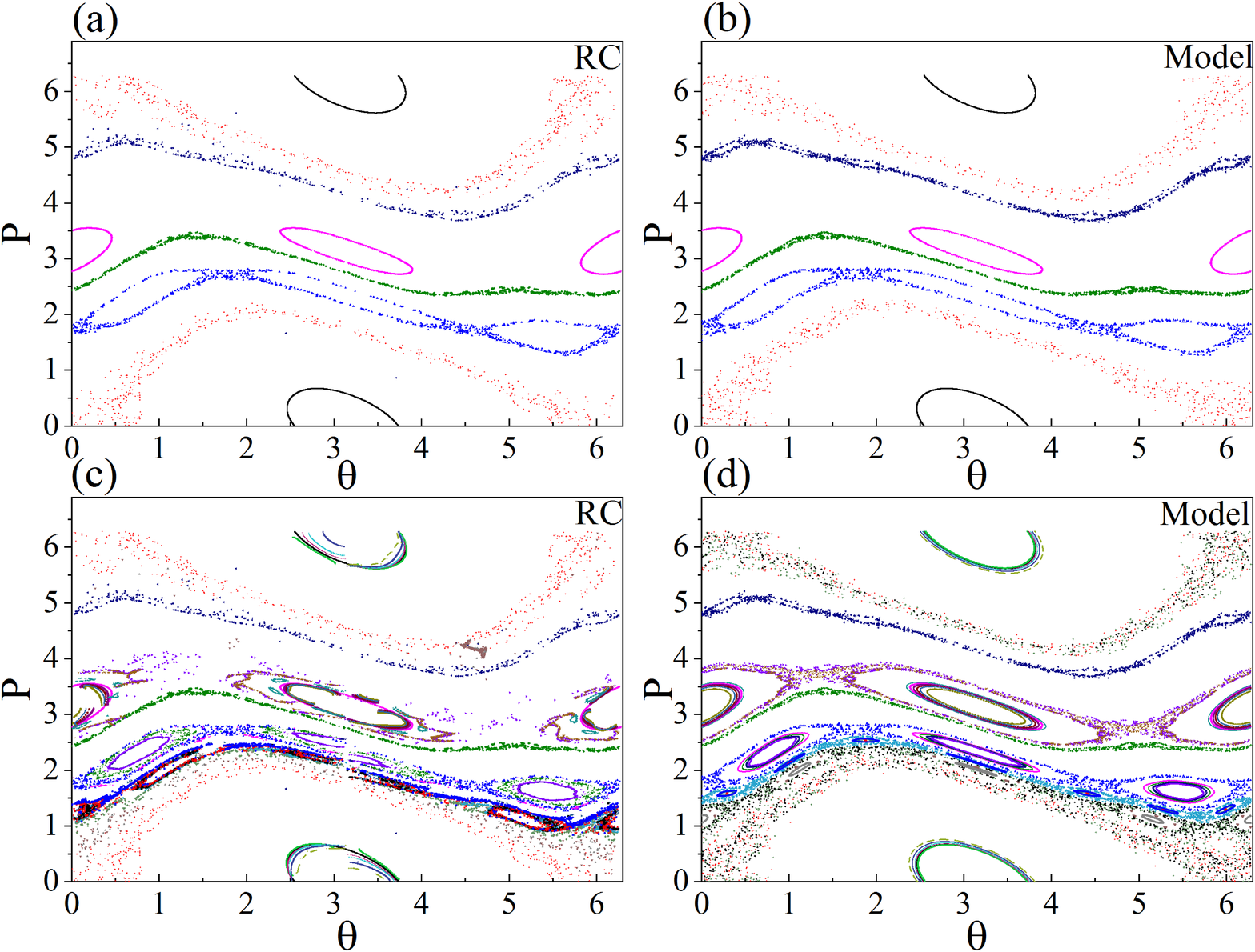}
\caption{Reconstructing the KAM diagram of the standard map under the kicking strength $K=1$. The training data is generated by $m=6$ training parameters. (a) The dynamics predicted by the machine for the training parameters. (b) The dynamics of the training parameters obtained from model simulations. (c) The diagram reconstructed by the machine, which includes the dynamics from the training parameters and $n=24$ dynamics from the additional control parameters. (d) For the same set of control parameters used in (c), the diagram obtained from model simulations.}
\label{fig7}
\end{figure*}

We next demonstrate the reconstruction of KAM dynamics diagram for a relatively strong kicking strength, $K=1$. For this kicking strength, many tori are destroyed and the diagram is mixed with quasi-periodic and chaotic motions. In this case, the training data are generated by $m=6$ training parameters randomly chosen from the range $[0,2\pi)$, $p_0=(0.58, 2.07, 2.19, 3.35,3.49,4.1)$ (three of them generate chaotic motions). The set of parameters for the reservoir are $(D_{r}, d, \rho, \alpha, \sigma, \lambda)=(1\times 10^3, 0.66, 0.77, 0.55, 3, 1\times10^{-9})$. The dynamics of the $6$ training parameters predicted by the machine are plotted in Fig.~\ref{fig7}(a). The corresponding results obtained from model simulations are plotted in Fig.~\ref{fig7}(b). We see that the climates of the training parameters are well replicated by the machine. By the trained RC, we replicate the climates for $n=24$ additional control parameters randomly chosen from the range $[0,2\pi)$. The replicated climates, together with the climates of the $6$ training parameters, are plotted in Fig.~\ref{fig7}(c). The corresponding results obtained from model simulations are plotted in Fig.~\ref{fig7}(d). We see that the KAM diagram predicted by the machine captures the main features of the diagram obtained from model simulations.  

\section{discussions and conclusion}\label{dc}

A few remarks on the performance of RC in learning Hamiltonian systems are in order. First, the purpose of the present work is to reconstruct the KAM dynamics diagram based on the time series acquired at a handful of training parameters, instead of a precise prediction on the state evolution of Hamiltonian systems. As such, the performance of RC is evaluated by the replicability of the statistical properties of the system dynamics, namely the climate, instead of the prediction horizon. Second, different from conventional RC in which the training and target systems are identical, in parameter-recognizant RC the target systems can be different from the training ones. From the standpoint of transfer learning~\cite{TL:SJP2010}, the fact that the KAM diagram can be reconstructed by the time series of a few training parameters implies the transferability of knowledge between different Hamiltonian motions. Finally, we would like to note that the performance of RC is dependent on the training data, including the number of the training parameters and the motions associated with these parameters. In general, the larger the number of the training parameters, the more accurate the reconstructed diagram. And, for the fixed number of training parameters, the more representative the motions of the training parameters, the more accurate the reconstructed diagram. For instance, for the example of standard map showing the mixed dynamics (see Fig.~\ref{fig7}), if all the $m=6$ training parameters are of quasi-periodic motions, the trained RC will be not able to replicate the chaotic dynamics and, as the consequence, the reconstructed diagram will be distinctly different from the true one. 

In machine learning of chaotic systems, a topic under active debate in literature is what the machine really learns from the data -- the mathematical expressions describing the given time series, the dynamics governing the system evolution, or the physical laws underlying the dataset and dynamics. For the purpose of state evolution prediction, the goal can be accomplished by fitting the time series with a mathematical expression, which, given the reservoir is complex enough, can be normally achieved~\cite{RC:Maass2002}. If this is the case, then the machine will be data-specific, e.g., the machine trained by periodic motions can not be used to replicate the dynamics of chaotic motions. Yet recent studies on transfer learning of chaotic systems show that in some circumstances knowledge can be transferred between systems of different motions, e.g., using the RC trained by periodic Logistic maps to replicate the climates of chaotic Logistic maps~\cite{KLW:2021}. The results of transfer learning suggest that it is the intrinsic dynamics that are leaned by the machine, but not the mathematical expressions describing the motions. Our current study of Hamiltonian systems suggests that the machine (specifically the parameter-aware RC) might learn something ``deeper" than mathematical expressions and dynamics, namely the physical laws. Physical laws in nature are characterized by symmetries and invariants, which are the essential rules defining the system dynamics and guiding the system evolutions. In learning physical systems, it is commonly believed that by incorporating the physical laws, the performance of the machines can be significantly improved, e.g., the development of HNN. Yet there are also studies showing that, just like the human brain, machine might be able to extract the physical laws and concepts from the data without any prior knowledge or assumptions about physics, kinematics, or geometry~\cite{MS:Science2009,WT:PRE2019,RI:PRL2020,YM:PRE2021}, saying, for example, Hamiltonians, Lagrangians, and other laws of geometric and momentum conservation. Our present work provides an additional evidence for the automated learning of physical laws from data by showing that the RC, probably the simplest recurrent neural network, is able to learn from the data the Hamiltonian mechanisms.

A brief discussion on the difference between HNN and RC is necessary. HNN is modified from the feedforward neural network by imposing the Hamiltonian mechanisms, in which the information is flowing from the input layer to the output layer in one-way fashion. As such, HNN can be essentially treated as a mapping function connecting the input and output vectors without intrinsic network dynamics. In particular, HNN has no memory about the previous input data and, without the input data, HNN can not generate the output data. In contrast, RC is a type of recurrent neural network and is essentially a complex dynamical system. For RC, the output not only depends on the input, but also is affected by the dynamical state of the reservoir. In particular, in the predicting phase the RC is evolving as an autonomous system, and is able to generate the output data continuously without the input data. Previous studies on machine learning of Hamiltonian systems are mainly based on HNN, in which the Hamiltonian mechanisms must be imposed on the algorithm~\cite{HNNs:Greydanus2019,HNNs:toth2019,HNNs:Bertalan2019,HNNs:Choudhary2020,HNNs:Mattheakis2020,HNNs:Cranmer2020,HNNs:Burby2021,HNNs:AC2021,HNNs:Dulberg2020,HNNs:Lai2021}. Our current study implies that for RC, the Hamiltonian mechanisms could be encoded in the reservoir in the training phase, and be used to reconstruct the KAM diagram in the predicting phase. In Ref.~\cite{HNNs:Choudhary2020}, it is speculated that ``an algorithm for training a recurrent neural network without a Hamiltonian loss function may be possible". Our study provides an affirmative answer to this speculation and, in the meantime, shows the transferability of knowledge in Hamiltonian systems.

Summarizing up, we have studied the learning of Hamiltonian systems by the technique of RC and found that, without prior knowledge of the Hamiltonian mechanisms, the trained RC is able to not only forecast the short-term evolution of the system state, but also replicate the long-term ergodic properties of the system dynamics. Furthermore, by the architecture of parameter-aware RC, we have demonstrated that based on the time series of a handful of training parameters, the trained RC is able to reconstruct the entire KAM dynamics diagram with a high precision. Although our studies are based on toy models, it is expected that the similar results can be also found in other Hamiltonian systems. The current study provides an alternative approach for learning Hamiltonian systems, and also sheds new lights onto the working mechanism of RC.

This work was supported by the National Natural Science Foundation of China under the Grant No.~11875182.

\end{document}